\documentclass[onecolumn]{aa} 
\usepackage{graphicx}
\usepackage{txfonts}

\newcommand{\myskip}[1]{}
\newcommand{\mBD}{$\mu$BD}
\newcommand{\mBDs}{$\mu$BDs}

\newcommand{\etal}{et al}
\newcommand{\MNRAS}{Mon. Not. Roy. Astr. Soc.}

\newcommand{\half}{\frac{1}{2}}

\renewcommand{\d}{{\rm d}}

\newcommand{\BEQ}{\begin{eqnarray}}
\newcommand{\EEQ}{\end{eqnarray}}
\newcommand{\BEA}{\begin{eqnarray}}
\newcommand{\EEA}{\end{eqnarray}}

\renewcommand{\d}{{\rm d}}

\begin{document}

\title{Do micro brown dwarf detections explain the galactic dark matter? } 

\subtitle{}

\author{
Theo M. Nieuwenhuizen\inst{1,2},
Rudolph E. Schild\inst{3} and
Carl H. Gibson \inst{4}}

\institute{
Institut de Physique Th\'eorique,
CEA Saclay,
F-91191 Gif-sur-Yvette Cedex,
France
\and
Institute for Theoretical Physics, University of Amsterdam, 
Science Park 904, P.O. Box 94485, 
1090 GL  Amsterdam, The Netherlands \email{t.m.nieuwenhuizen@uva.nl}
\and
Harvard-Smithsonian Center for Astrophysics, 60 Garden Street, Cambridge, MA02138,
USA. E-mail: \email{rschild@cfa.harvard.edu} 
\and
Mech. and Aerospace. Eng. \& Scripps Inst. of Oceanography. Depts., UCSD, La Jolla, CA
92093, USA. E-mail: \email{cgibson@ucsd.edu} }

\abstract
{The baryonic dark matter dominating the structures of galaxies is widely considered as
mysterious, but hints for it have been in fact detected in several astronomical observations at
optical, infrared, and radio wavelengths. We call attention to the nature of galaxy merging, 
the observed rapid microlensing of a quasar, the detection of ``cometary knots'' in planetary nebulae, 
and the Lyman-alpha clouds as optical phenomena revealing the compact objects. Radio observations 
of ``extreme scattering events'' and ``parabolic arcs'' 
and microwave observations of ``cold dust cirrus'' clouds are observed at 15 -- 20 K temperatures
are till now not considered in a unifying picture.}
{The theory of gravitational hydrodynamics predicts galactic dark matter arises from Jeans clusters
that are made up of almost a trillion micro brown dwarfs (\mBDs) of earth weight. 
It is intended to explain the aforementioned anomalous observations and to make predictions within this framework.}
{We employ analytical isothermal modeling to estimate various effects.}
{Estimates of their total number show that they comprise enough mass to constitute the missing
baryonic matter.  Mysterious radio events are explained by \mBD \ pair merging in the Galaxy. 
The  ``dust'' temperature of cold galaxy halos arises from a thermostat setting due to
a slow release of latent heat at the 14 K gas to solid transition at the \mBD \  surface. 
The proportionality of the central black hole mass of a galaxy and its number of globular clusters 
is explained. The visibility of an early galaxy at redshift 8.6 is obvious with most hydrogen locked up
in \mBDs.}
{Numerical simulations of various steps would further test the approach. It looks promising to redo
MACHO searches against the Magellanic clouds.}



\authorrunning{Nieuwenhuizen, Schild \& Gibson}

\maketitle


\section{Introduction}

In 1931 Jan Oort realized that our Galaxy must contain invisible matter (\cite{Oort}).  Fritz
Zwicky soon concluded the same for clusters of galaxies, and called it dark matter (DM) ~(\cite{Zwicky}).
The current paradigm of Cold Dark Matter (CDM, or $\Lambda$CDM including the cosmological constant)
 assumes it to arise from a heavy elementary particle with a small interaction cross-section. 
In the transition of the plasma of protons,
electrons and He-ions to gas, the photons decouple and the baryons are said to condense on presumed
pre-existing CDM condensations. However, despite many searches, this CDM particle has
not been found ~(\cite{Xenon100,CDMS2}) and this bottom-up scenario is stressed by numerous
observations at its upper and lower size scales. On the largest scales, cosmic void structures
are found on much larger size scales than predicted from simulations, whereas on the smallest 
scales, the many subhalos predicted to surround our Galaxy are not observed ~(\cite{Diemand2008}).
Moreover, the voids are emptier than simulated. And, the predicted  ``dark age" before the
formation of stars and galaxies is coming more and more under stress with new observations
of early structures arising basically every month ~(\cite{Oesch,Bouwens2009}). 
Despite careful re-analysis the $^7$Li abundance  
remains a factor 4--5 below the WMAP $\Lambda$CDM prediction~(\cite{Cyburt2008}).
Neither does $\Lambda$CDM offer an explanation for the axis of evil ~(\cite{axisofevil}),
the dark flow ~(\cite{darkflows}), chain galaxies ~(\cite{chaingalaxies}) or the bullet cluster
(\cite{LeeKomatsu2010}), 
while correlations between galaxy parameters contradict the $\Lambda$CDM hierarchical clustering 
~(\cite{Disney2008}).
The structure of the Local Group~(\cite{Kroupa2010}) and globular star clusters in the Galaxy
~(\cite{Conroy2010}) are incompatible with CDM.

We shall follow an approach with two types of dark matter, ``Oort" DM in galaxies,
composed of baryons, and ``Zwicky" DM in galaxy clusters, the true DM.
Gravitational hydrodynamics (GHD) theory of structure formation assumes no new particle 
but only stresses the nonlinearity of hydrodynamics. It involves a top-down 
scenario starting with structure formation in the plasma at redshift $z = 5100$, well before the   
decoupling of matter and radiation at z = 1100 ~(\cite{Gibson1996,NGS2009}).
The viscosity in the quark-gluon plasma is very small, so it leads to strong turbulence.
This gets dissolved, but as is known from turbulence in air fares, pockets of turbulence survive and
trigger nonlinearities at the moment the viscous length enters the horizon at redshift $z=5100$.
This viscous instability mechanism creates voids, which have expanded to ca 38 Mpc now, a fair estimate for the observed
typical cosmic voids ~(\cite{GellerHuchraVoids}). Just after the decoupling, the Jeans 1902 mechanism says
that all gas fragments in Jeans clumps of ca 6$\cdot10^5$ solar masses.  
These clumps also fragment, again due to viscosity but now of the gas, in 
some $2\cdot 10^{11}$ micro brown dwarfs (\mBDs, muBDs) 
of earth mass, sometimes called MACHOs (Massive Astrophysical Compact Halo Objects). 
This process turns the Jeans clumps into Jeans clusters (JCs) of \mBDs.
The first stars occur by coagulation of \mBDs, on gravitational free-fall time scales without a dark period. 
A fraction of Jeans clusters turned into old globular star clusters. Others got disrupted and became material 
for normal stars. But most of them still exist, though cold, and constitute the galactic dark matter. 
These dark JCs have an isothermal distribution, noticed in lensing galaxies
 ~(\cite{Hubble,isothermal}), which induces the flattening of rotation curves, 
 while their baryonic nature allows to explain the Tully-Fisher and Faber-Jackson relations ~(\cite{NGS2009}).

Nuclear synthesis attributes enough matter for the \mBDs. About 4.5\% of the critical
density of the Universe is baryonic. At best 0.5 \% is luminous, and some 2--3\% is observed in X-ray
gas.  So, the missing dark baryons may indeed be locked up in \mBDs.

Questions about the nature of the baryonic dark matter (BDM) have largely been
overshadowed in the past decade by even deeper questions about the non-baryonic dark
matter. GHD just assumes that it is free streaming at the decoupling~(\cite{NGS2009}).  One  of us in his modeling of the lensing properties of a galaxy cluster ~(\cite{NEPL2009}) confirmed that it should be 
related to cluster DM. Dark matter is described by an ideal gas of isothermal
fermions, and galaxies and X-ray gas by standard isothermal models. This yields a DM mass of 1-2
eV, not the heavy mass of the hypothetical CDM particle. The best candidate is the neutrino
with mass of 1.5 eV; a prediction that will be tested in the 2015 KATRIN tritium decay
search~(\cite{Weinheimer2009}). 
If also the right-handed (sterile) neutrinos were created in the early Universe, there
should exist a 20\% population of 1.5 eV neutrino hot dark matter, clumped at the scale
of galaxy clusters and groups~(\cite{NEPL2009}).

Returning to the problem of the more familiar galactic DM just consisting of baryons, it has probably been 
detected already in various observations but has not been considered in a single framework. 
The nature of the BDM seems obscure because nature places severe constraints, 
so it has thus far eluded detection. The BDM must be dominated by hydrogen in
some form.  Hydrogen can exist in its atomic form, as H$_2$ gas, or in ionized form, H$^+$. 
In gaseous form as H-clouds its strong signatures as
radio (21 cm) and optical Lyman and Balmer alpha emission lines should have been observed
in some kind of galaxy or other by now, creating a serious puzzle, that we solve here.

A second important impediment faced by the idea that the BDM is sequestered away
in condensed objects is the formation question. It has long been concluded that hydrogen
clouds cannot collapse to form planetary or solar mass objects because the hydrogen cannot
cool quickly, and pressure forces will quickly restore the cloud. However these arguments all
involve a simplifying assumption that the process can be linearized, whereas GHD 
starts from H condensed in \mBDs.
We now discuss that various types of observations support the GHD picture as the CDM theory flounders.

\section{Jeans clusters visible in galaxy merging}

The GHD formation theory predicts some $3\cdot 10^{17}$ \mBDs \ of earth mass per galaxy 
grouped in some 1.7 million Jeans clusters, each with mass $6\cdot 10^5 M_\odot$, that constitute the 
galactic DM halos~ (\cite{Gibson1996}).  \mBD \ merging has led to heavier objects like planets and stars.

As exposed in Fig. 1 of the Tadpole galaxies, galaxy merging turns the dark halo JCs into young (3-10 Myr)
globular clusters (YGCs) along the merging path~(\cite{GibsonSchild2002,NGS2009}). 
The age distribution has been determined, with typical value of 10 Myr, but many are younger~(\cite{Fall2005}).
Similar behaviors are observed in e. g. the Mice and Antennae galaxy mergers.
The most luminous of the ``knots" in Fig. 1 has an age of 4--5 Myr and estimated mass  
$6.6\cdot10^5M_\odot$~(\cite{Tran2003}), reminiscent of a JC.
Observations generally reveal that the bright blue clusters are much younger (a few Myr) than the dynamical 
age of the tail (ca 150 Myr), providing strong evidence that star formation occurs in situ, long after the tail was 
formed ~(\cite{Tran2003}). 
GHD asserts that the cold \mBDs \ along the merging path get warmed by tidal forces, 
which makes them expand and coalesce at suitable moments into young stars grouped in YGCs.
The $\Lambda$CDM explanation as disrupted tidal tails ~(\cite{Duc}) suffers from predicted but unobserved old stars.

Estimating the amount of galactic dark matter from fig. 1, the tail begins at 420,000 
lyr from the center where it exhibits the sharp boundary of the dark matter halo. If there is spherical
symmetry, the enclosed mass is $2.4\cdot10^{12}M_\odot$, a reasonable estimate.
The opening width of $1500$ lyr across remains fairly constant along 
the trail, as expected for an induced local heating effect.
From an isothermal model the number of JCs in the wake can then be estimated as 5900,
which within a factor of a few coincides with the number of bright spots in Fig. 1.
Indeed, some 11,000 were analyzed in Ref. ~(\cite{Fall2005}).

From these features one can already conclude  
that the galactic dark matter likely has the structure predicted by GHD.

\begin{figure}
\centering
\includegraphics[width=10cm]{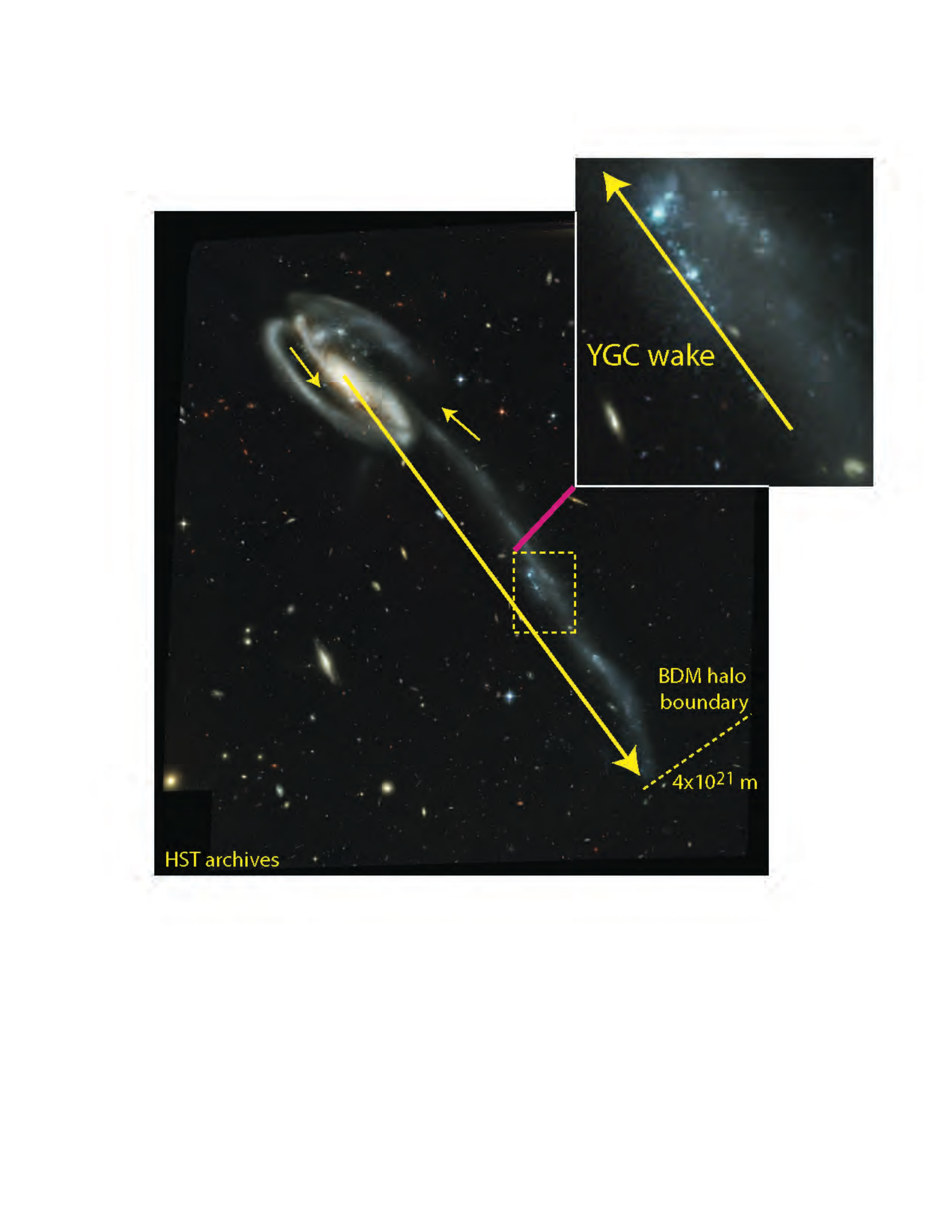}
\caption{As the Tadpole galaxies merge, they heat up their dark baryonic halos of Jeans clusters 
of cold micro brown dwarfs. The latter expand, coalesce and form new stars, turning the Jeans 
clusters into young globular clusters (YGC), visible as a trail of many hundreds of bright dots as the merging galaxy fragments pass through the main galaxy disk.}
\end{figure}

\section{Quasar microlensing brightness fluctuations}

Quasar micolensing has probably to date given the most compelling evidence that the
BDM is contained in a network of planetary mass bodies with sufficient
number to account for the entire baryonic dark matter. The first detection of this was made
with brightness monitoring of the Q0957+561 A, B gravitationally lensed quasar, with two
images A, B separated by 6 arcsec on the plane of the sky. 15 years of brightness
monitoring revealed that a microlensing signal of amplitude 1\% and time scale 1 day
(observer's clock) was apparent in the data ~(\cite{Schild96}). Because the signal was consistently
 observed for long periods of monitoring on a time scale of the monitoring frequency, 1 day, it was
concluded that the microlensing originated in a population of compact objects having masses
averaging $10^{-5.5} M_\odot$ and with other events up to Jupiter mass also seen.

Subsequent to this discovery, re-analysis of the quasar data ~(\cite{Pelt1998}) confirmed that the rapid
fluctuations were not intrinsic to the quasar and therefore presumably microlensing.
Furthermore the signal had an equal positive and negative network of peaks, as demonstrated
from wavelet analysis ~(\cite{Schild1999}). This is understood to originate from a high optical depth and
therefore is a signature that the \mBD \ population observed in the lens galaxy must be the
baryonic dark matter. This conclusion follows from a deep understanding of microlensing 
~(\cite{Schneider1992}). In
general, the shear associated with gravitational lensing (macro, milli, micro, nano) causes a
brightening event with a cusp-shaped profile. But in the presence of a macro-shear generated
by the Q0957 lens galaxy G1, a micro-shear caused by a \mBD \ in the lens galaxy would
possibly further amplify the macro-shear or potentially de-magnify it on the micro-arcsec
scale. Only in the special circumstance that the micro-shear originates in a population with
unit optical depth to microlensing will the probability of
micro-magnification and micro-demagnification
be equal (p. 343 and Fig. 11.8 of Ref ~(\cite{Schneider1992}) leaving a brightness profile with
equal brightening and fading events on the time scale specific to the microlensing compact
object, as simulated by (\cite{SchildVakulik2003}).
Thus, the observed signature found to have approximately equal positive and negative
events ~(\cite{Schild1999}) uniquely indicates microlensing by a population at unit optical depth. Since the
optical depth of image A is known to be 0.35 and of image B is 1.35, as determined by a
model of the overall lensing that produces the double image, and since the lens galaxy G1
must plausibly be dominated by its baryonic dark matter, it may be concluded that the
baryonic dark matter is detected as a population of \mBD \ objects in the lens galaxy G1.

In an isothermal modeling, the average number of JCs along the quasar 
sightline A is 0.002 and for B 0.01. These numbers are close enough to unity to conclude that 
along each sight line there lies exactly one JC.  The estimated number of \mBDs \ along the 
total light path then comes out as 18,000, confirming the observed multiple lensing events.
The typical induced optical depth is 58, but the observed lower values
occur if the light path traverses near the border of the JCs.

We are left with the extrapolation from theory and observation that the entire baryonic
dark matter occurs in the form of a population of \mBDs \ seen in
microlensing of quasars by the grainy distribution of matter in the lens galaxy. The same
population has been confirmed in several other lens systems 
~(\cite{Burud2000,Burud2002,Paraficz2006}), with microlensing mass estimates in the range 
$10^{-5.5}$ to $10^{-3}M_\odot $, that is, from individual \mBDs \ to merged ones.

\section{Cometary knots in planetary nebulae}

The \mBDs \ are directly seen in the Helix and other planetary nebulae, see Fig. 2.
Their infrared emission allows 20,000--40,000 of them to be counted in Spitzer  
~(\cite{SpitzerHelix}).
They are found distributed in a band or shell at a central distance comparable to the
Solar Oort cloud, $7\cdot10^{15}$ m.
Because they are objects  resolved at optical and infrared wavelengths, their sizes and
structures may be estimated. Since they are radio CO emission sources, their masses may be
estimated from their total CO emission and a normal CO/H emission ratio~(\cite{Huggins 2006}). 
They have a cometary shape which allows their masses to also be estimated from an ablation theory
applied to their outer atmospheres ~(\cite{Meaburn1998}). Both methods yield mass estimates between
$10^{-6}$ and $10^{-5} M_\odot$. The current literature often explains them as
Rayleigh-Taylor instabilities in
the outflowing nebular gas. However, RT instabilities would not cause strong
density contrasts in the ambient gas.  Density contrasts of only 4-6 would be expected from the Rankine-Hugoniot supersonic jump condition, whereas density contrasts of 1000 are observed from direct mass estimates 
~(\cite{Meaburn1998}).
Attempts to measure their proper motions by comparing HST images at different epochs have
been carefully done to define whether they have expanding or static
distributions ~(\cite{ODell1996, BurkertODell1998}), but the results have been inconclusive. The gas
outflow velocity of ~(\cite{Meaburn2010}) does not refer to the proper
motion measured expansion of the system of knots.

\begin{figure}
\includegraphics[width=9cm]{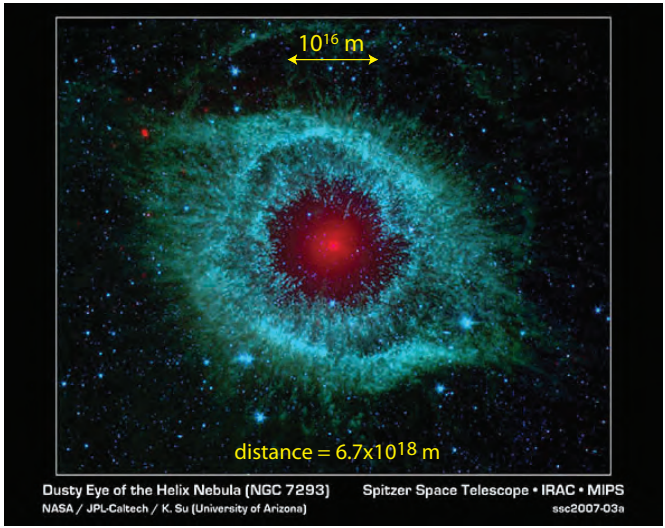}
\vspace{-3.5cm}
\caption{Infrared observation by Spitzer of the Helix planetary nebula. 
The white dwarf in the center is surrounded by an Oort cavity around which some 20,000--40,000 
micro brown dwarfs are observed in a band perpendicular to the line-of-sight. 
They are visible due to heating by the plasma jet of the white dwarf.}
\end{figure}

\section{On direct MACHO searches}

With gas clouds apparently absent, the community turned, in the 1990s, to searches for condensed
hydrogen-dominated objects and initiated microlensing searches for them.
The MACHO ~(\cite{Macho1,Alcock2000,Alcock2001}), EROS ~(\cite{Aubourg1993,Eros1,Eros2}) and 
OGLE ~(\cite{Ogle,Ogle2,Ogle3}) programs all found microlensing events in front of the 
Large and Small Magellanic Clouds (MCs) and the galactic center.
Early claims that the few detections implied a population of white dwarfs were  
not substantiated because the population of progenitor stars could not be identified. 
Most observed objects are white dwarfs, Jupiters or Neptunes, probably inside the MCs
and galactic center.

From GHD one expects all galactic DM  of earth mass \mBDs.
This case is just allowed by the MACHO collaboration~(\cite{Macho1}),  but ruled out by 
EROS-1~(\cite{Renault1998}). However, for observations against the Magellanic clouds, the estimate for
the radius of the \mBD \  of $2.6\cdot 10^{11}$ cm coincides with its typical Einstein radius, 
leading to the microlensing unfriendly finite-source finite-lens situation~(\cite{Agol2002,Lee2010}). 
Obscuration by \mBDs \ and refraction by their atmospheres may have hindered the observation 
of Macho events.

For the  \mBDs \ now predicted to be grouped in JCs, blind searching is not optimal. 
The JCs in front of the Magellanic clouds can be identied from WMAP, Herschel or Planck imaging data. See them also in the DIRBE-COBE image of fig 3. They should be a fertile ground for MACHO searches, 
but the observation cadence must be greatly increased because of the low mass of the \mBDs. 
For a terrestrial mass MACHO passing in front of a small star the total event duration is approximately 3.5 hrs. For \mBD \ microlenses clustered with properties measured for ordinary globular clusters, 
some 4,100 events should be visible per year by a dedicated spacecraft that continuously monitors 
the stars behind any of the 3400 JCs in front of the MCs. 
The number would increase if a clump of clusters as described below lies in front of the LMC.
There should also be multiple events on the same star.

\section{Halo temperatures and cirrus clouds}

If most of the baryonic dark matter is sequestered away from view in \mBDs \
its thermal emission should still be seen as structured far-IR or sub-mm emission. 
The \mBDs \ would have been warmer than ambient gas at time of \mBD \ formation, 
and with their gravitationally heated cores they would be
radiating heat away to the universe. Initially this would have cooled the
objects significantly on time scales of 1 million years, since it has
already been noticed that they are observable as T-dwarfs only in the
youngest star forming regions~(\cite{Marsh2010}).
Thereafter they would have continued to cool in the expanding and cooling
universe, but with large cooling blanketing atmospheres they should then have
cooled slowly. 

Because hydrogen has a triple point at 13.8 K, higher than the 2.725 K 
cosmic microwave background, phase transitions with significant heats of
condensation and fusion have the potential to act as a thermostat
to set the temperature of the baryonic dark matter, thus
explaining the cool galactic dust temperatures of  $<20$K 
~(\cite{Radovich2001})  and 14--16K and 15--25K~(\cite{Liu2010}) as galaxy haloes.
Indeed, cooling atomic hydrogen at low density and pressure passes slowly through its
triple point, so the \mBD \ population should have dominant thermal emission at 
or slightly above this temperature.
The wavelength of peak emission would be 210 microns, or 0.21 mm. 
A detection of this thermal signature has also been made in a careful
comparison of IRAF, DIRBE, BOOMERANG, and WMAP images for a Halo region of
our Galaxy. These show ~(\cite{Veneziani2010}) that the observed structure
radiates at approximately 16K and has its peak radiation in the 220 micron
DIRBE band. In fig. 3 we show one panel of fig. 2 of (\cite{Veneziani2010}),
which is now understood to be an image of the baryonic dark matter, 
seen in its emission peak wavelength.
Although the far-IR emission of the Halo has been described as ``cold dust cirrus" 
we find from direct imaging a clumped distribution, even with nested clumps, 
which suggests an interpretation of it as a clumpy distribution of JCs.

This view is supported by the amplitude of the signal. The standard Planck intensity 
at the DIRBE frequency of 1.25 THz caused by a thermal \mBD \ atmosphere at 15 K is 54 GJy. 
With an \mBD \ radius of $2.6\cdot 10^{9}$ m and the typical distance to the next
JC  equal to $8.9\cdot 10^{18}$ m, the Planck intensity is attenuated by the square of their ratio 
to become 0.9 kJy for the whole JC, about the highest observed value in ~(\cite{Veneziani2010}).
Moreover, that nearest JC would have an angular diameter of $0.6^\circ$, slightly less than
the largest structure of Fig. 3. Next, the intensity per solid angle can be estimated 
as 12 MJy/sr, mildly overestimating the maximal scale of 2 MJy/sr of several objects in Fig. 3. 
We can also estimate the number of structures in Fig. 3. From an isothermal model we
predict about 67 structures larger than 0.25$^\circ$ in diameter, which is a fair estimate.

With reasonable estimates for their intensity, angular width and number, it is likely that the 
concentrations in Fig. 3 arise from the JCs that constitute the full Galactic dark matter.
Non-baryonic dark matter is relevant only on galaxy cluster scales.

A further aspect is the size of the \mBD \ atmospheres. For redshift $z>4$ the ambient temperature is 
above the triple point and atmospheres are large. After that, liquid drops form in the outer atmospheres 
which rain onto the center. This will lead to a sizable shrinking of the atmospheres,
and enhanced transmissivity for radiation, an effect similar to the reionization of the inter cluster gas.
The latter may be caused by neutrino condensation on the cluster~(\cite{NEPL2009}). 

\begin{figure}
\includegraphics[width=10cm] 
{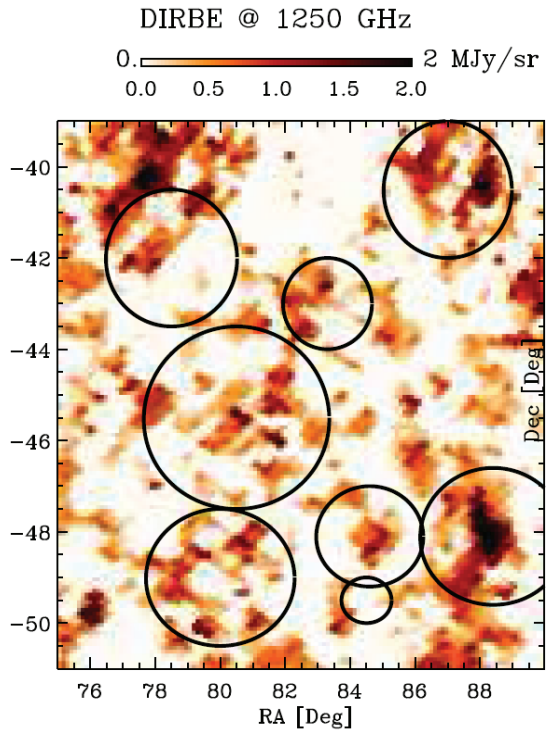}
\vspace{-1.5cm}
\caption{The DIRBE map of a $15\times12$ degree region of the Halo, taken from Fig. 2 
of ~(\cite{Veneziani2010}), with black circles denoting regions focused on in that work.
Emission knots at the 240 micron peak of the thermal emission 
at the 15 K freezing-boiling point of hydrogen are
here explained as a few hundred optically dark Jeans clusters 
at various distances aggregated in clumps.} 
\end{figure}

\section{Radio events with permanent sources}

Extreme Scattering Events were discovered as radio brightness anomalies in quasars.
An event was characterized by 30\% amplitude fluctuations with a cusp-profiled signature that
is somewhat frequency dependent. This allowed them to be understood as a refraction event
caused by a centrally condensed gaseous object passing in front of the quasar with a
transverse velocity of a few hundred km s$^{-1}$, the radio analogue of the above quasar
brightness fluctuations. A physical model was quickly developed, which
included refraction by an ionized outer atmosphere acting as a negative lens that produces the
observed pattern of brightness cusps ~(\cite{WalkerWardle1998}). The objects were described as self-gravitating clouds
with approximately spherical symmetry. In the refraction model, an electron column
density of 10$^{20}$m$^{-2}$ is determined. The atmospheres of the objects have a size of 1 AU 
($1.5 \cdot 10^{13}$ cm),  and must
be gravitationally bound or else confined by an ambient pressure. We notice a similarity to
the properties of the \mBD \ discovered in quasar microlensing. The ESE clouds would be self-gravitating,
have mass less than 10$^{-3}M_\odot$, have a size of 10$^{14.5}$ cm, are a Halo
population, and have typical galactic speeds. Not surprisingly, it is independently estimated
that the observed objects may relate to a significant fraction of the galactic DM ~(\cite{WalkerWardle1998}).


Pulsar anomalies also suggest the existence of compact objects in the Halo of our Galaxy
~(\cite{Hill2005}),  as already inferred from quasar anomalous brightening~(\cite{WalkerWardle1998}). 
Estimated sizes are of order 0.2 AU ($3\cdot10^{10}$ m).


A probably related phenomenon is the crescent shaped events seen in the timing data
in pulsars, since like the Extreme Scattering Events in quasar brightness curves, the pulsar
anomalies are attributed to a refraction phenomenon due to condensed objects along the sight
lines. It is inferred that the refraction effect produces multiple images of
the pulsar radio signal, and interference of the images causes the observed phenomena.
The scattering medium is estimated to have structure on 1 AU for a 10$^3$ interstellar density
enhancement~(\cite{Walker2007}). This corresponds to a radio scattering atmosphere of 
$10^{16}$ kg, as expected well below the GHD prediction of an earth mass for the entire object.

\section{Mysterious radio events as \mBD \ merging}

Recently, a mysterious class of ``long duration radio transients'' has been observed,
lasting more than 30 minutes but less than several days, that have neither a counterpart in the 
near-infrared, visible or X-ray spectrum nor a quiescent radio state.
The event rate is very large, $\sim10^3$deg$^{-2}$yr$^{-1}$ and they can be bright ($>1$Jy).
They were attributed to old galactic neutron stars~(\cite{Ofek2010}, but this would result in more
luminosity for distant galaxies than observed.

Within GHD it is natural to connect these events to \mBD \ mergings in the JCs that constitute the BDM 
halo of the Galaxy. Let us present a statistical estimate for the event frequency.
{}For the typical duration we take $t_{ev}=1$ day, so the effective \mBD \  radius 
is estimated as $R_{bd} = v_{bd} \,t_{ev}=2.6\cdot 10^{11}$ cm.
The typical JC has $(R_{jc}/R_{bd})^3$ cells of linear size $R_{bd}$.
At a given moment in time the average number of  \mBD \  pairs
that occupy the same cell in a given JC is $0.5$. In the Galaxy the number of Jeans clusters is $1.7\cdot 10^6$.
The resulting rate of merging events is $7400$/deg$^2$yr,  modestly overestimating the observed value.

The amount of gravitational energy available in the merging process is large, 
of order $GM_{bd}^2/R_{bd}$, which means some $10^5$ J/kg.
Assuming this to be emitted at distance of 3 kpc during the event time of 1 day in a frequency
band of 10 GHz gives a bright radio event of 1 Jy; such events are observed,
see table 2 of ~(\cite{Ofek2010}).  Ofek et al. also discuss scenarios for the 
emission in radio such as synchrotron radiation.

\section{Direct detections in nearby star forming clouds}

A star formation region in NGC 5253 at distance of a few Mly has been observed in radio, 
corresponding to some 1200 O7 stars in a parsec-sized central region ~(\cite{Turner2004}). 
This is understood to represent the formation of a young globular cluster with some 100-200 times 
more stars in total, out of supposed molecular clouds.
The recombination linewidth of 75 km/s has been interpreted to give a mass of 4-6 $10^5M_\odot$
~(\cite{Turner2004}). In our picture this central region is a Jeans cluster with the right size and mass, 
which is on its way to turn into a young globular cluster by star formation out of its \mBDs. 
The region as a whole, with some 7000 O7 stars within 30 pc, can be explained as 
an aggregation of JCs.

Recently a 2-3 Jupiter mass object has been discovered in the $\rho$ Ophiuchus star forming 
region ~(\cite{Marsh2010}). Estimated properties were a T spectral type, a surface temperature 
near 1400 degrees K, and an age of one million years. At its inferred distance of 100 pc it is
presumed to be a member of the $\rho$ Ophiuchus cloud region because of its heavy visual
absorption. Since active star formation in the region is underway, it is inferred that the object
has formed in the standard cloud collapse scenario, and represents an object near the bottom
of the stellar mass function. It cools quickly (in about a million years), so that it may
represent a population of planet-mass MACHO objects. With reasonable assumptions about
their luminous lifetimes and number, they have been inferred to be a cosmologically
significant population ~(\cite{Marsh2010}).  In GHD one sees the Jupiters as clumps of \mBDs,
that may go on growing to a star. Hence this observation hints at the population of \mBDs.

This discovery is discussed ~(\cite{Marsh2010}) in the context of the many similar objects found 
in the most active star forming regions in Orion ~(\cite{ZapateroOsorio2000,Lucas2006}). 
All these authors recall that such objects would not form in the framework of the standard 
linearized theory of star formation, and appeal to fragmentation models ~(\cite{Whitworth2006}) 
which, however, seem incompatible with the extreme binarity seen in the statistics of double stars 
and our solar system's Kuiper Belt Objects.

In Table 1 we show multiple estimates of the same physical parameter where they
have been independently estimated in different research programs related to the \mBDs.
The masses are one earth mass and up.
Depending on their temperature they may have very different sizes and densities, 
since they are H-He gas clouds. 

\begin{center}     
\noindent     
\begin{tabular}{||l|r|r|r|r|r|||} 
 \hline \hline     
Estimator&microlensing&\hspace{4mm}ESE\hspace{4mm}&Cometary knots&T dwarfs\\     \hline\hline
Log Mass ($M_\odot$ ) &--5.5& $< -3$&  --5.3 (--5.2) &--2.6\\ \hline
Log radius (cm) &--& 14.5&  15.6&  9.9\\ \hline
Log density (cm$^{-3}$)& -- &12& 6,(5.9), (5.6)& 24.2\\ \hline
Velocity (km s$^{-1}$) &600& 500& --& -- \\ \hline \hline
Cosmological Signf. &yes& yes& -- & yes \\  \hline     
\hline   
\end{tabular}     
\end{center}     
\vspace{6mm}   

Table 1. Measured and inferred properties of micro brown dwarfs from various observations.
ESE = Extreme Scattering Events.

\section{Inferred source of Lyman-alpha clouds}

In quasar spectra a population of thousands of hydrogen-dominated clouds is seen
causing weak absorption lines called the Lyman-alpha forest. Their properties have been
reviewed ~(\cite{Rauch1998}). Since they are detected as redshifts of the 1216 \AA \ 
Ly-$\alpha$ \ line and are sometimes accompanied by higher level Lyman lines, their identification is secure. 
In a typical quasar spectrum, one or two stronger and damped lines are also seen. In these cases 
there are weak metallic absorption lines at the same redshift. Because they are found along sight
lines to cosmically distant quasars, much is known about their distribution from their redshifts. 
They are known to be relatively uniformly distributed in redshift,
with, however, mild clumping at distance/size scales of 20 Mpc,  showing the cosmic
pattern of voids on scales of 30 - 130 Mpc. The depth of cloud absorption lines and their
numbers increase with redshift $z$. Perhaps their most interesting property from a structural
point of view is the velocity width parameter measured for each line. The average value is
approximately 30 km/s, which suggests that as clouds they should dissipate on a time scales
of several years. 
Their observed permanence was first presumed to imply that the clouds were
confined by a hot inter-cluster medium, which was searched for but not found. This has left
the subject with no explanation for their structure and common existence. 

We propose that the clouds are in fact outer atmospheres of the \mBD \  population in JCs detected 
in other observations as inferred above. 
JCs intersecting by the quasar sight line lie in a cylinder, so their mass can be estimated  as 
$0.02\rho_c\cdot\pi R_{jc}^2\cdot c/H_0$, where the factors describe the average cosmic baryon 
density not in X-ray gas, the JC surface area and the typical quasar distance, respectively. 
Though this corresponds to 0.0001 JCs only, a few thousand may actually occur
because matter is clumped in filaments between voids, where the baryonic density
may be 5-10 times the mean galactic mass density, i.e. about $10^7\rho_c$. 
In a JC located close to the quasar,
about 900  \mBDs \  will lie in front of it,  which by their random motion at 30 km/s cause
an absorption line of width 30 km/s. This number becomes lower for JCs closer to us, 
because the opening angle of the quasar covers less of their surface.
At higher redshift the \mBD \ atmosphere will be warmer and larger, which explains the increase
of Ly$_\alpha$ absorption with $z$~(\cite{Rauch1998}).

We attribute the broad absorption lines corresponding to b = 200 km s$^{-1}$  to a set of JCs along the 
line of sight with this velocity dispersion typical for objects in galaxies, while the involved mass 
could then be large enough to expose the metal lines.
With most JCs located in galaxies, the typical number of JCs per galaxy pierced by the light path 
of the quasar is $0.001$, which explains that per few thousand narrow absorption lines there will
typically be a few broad ones.

\section{Iron planet cores}

A perplexing problem is the origin of the magnetic iron cores of the Earth, Mercury and
Neptune. It is well known that most of the metallic atoms in the universe are seen in
meteoritic dust as oxides of iron, silicon etc. So it is difficult to imagine how the iron was
reduced to metallic form underneath an ocean of water. However, primordially formed \mBDs  \  offer
a direct scenario. From their periodic mergings to form larger planets and eventually
stars, they have well-mixed and massive gas and liquid-solid hydrogen layers
with wide temperature ranges that gravitationally sweep up and reprocess
  iron and nickel oxide supernova II star dust from the interstellar 
  medium (\cite{schilddekker}. Metal oxide dust grains meteor into the hydrogen
layers where reaction kinetics and hot mixing mergers reduce iron, nickel and other oxides to metals
 in solid and liquid forms that sink to form dense metal core layers under oxide layers
 (rocks, water) with cold outer hydrogen layers. 
Planets near the sun lose most of their hydrogen atmospheres in their
pre-stellar accretion discs and by the solar wind.

\section{Black holes, young stars and globular clusters}

Our conclusion that the long sought baryonic dark matter has been observed already as micro 
brown dwarfs of planetary mass has implications for the formation process of planets and stars. 
The observation of young stars (age $\sim$1 Myr) in a disk near the black hole (BH) in the
center of our Galaxy known as Sag A$^\ast$, is difficult to explain and termed the 
``paradox of youth" ~(\cite{Ghez}.
A new scenario is offered by GHD: in a Jeans cluster passing close by the BH the \mBDs \ 
were heated by the strong tidal forces, they expanded and coagulated in situ into new stars; 
this JC got disrupted and its stars ended in a plane.

Tidal forces by heavy central BHs may generally have a strong impact on nearby passing JCs.
Let us suppose that JCs have an appreciable chance to be transformed into a globular star cluster ({\it gc}) 
when their gravitational energy $GM_{BH}M_{jc}/R$ exceeds a certain bound $E_c$, which happens when 
they pass within a distance $R_\ast\sim M_{BH}$. In an isothermal model their number $N(R_\ast)$ 
is proportional to $R_\ast$,  explaining that $N_{\it gc}\sim N(R_\ast)$ is proportional to $M_{BH}$, 
in accordance with the observation $M_{BH}\sim N_{\it gc}^{1.11\pm0.04}$ ~(\cite{BurkertTremaine2010}). 

A population of extended star clusters in the discs of two nearby galaxies is called faint fuzzies  
~(\cite{Burkert2005}), but their determined properties are as expected for a recently formed stars 
in a population of JCs.

\section{Paradox on the visibility of early galaxies}

After the submission of the original manuscript the galaxy UDFy-38135539 was established to have 
redshift $z=8.55$, so we see light that it emitted 600 million years after the big bang~(\cite{Lehnert2010}).
This is believed to be well before the reionization era, so it is a question how the opaque hydrogen gas
could allow the visibility. The authors say in the abstract: 
{\it We find that this single source is unlikely to provide enough photons to ionize the volume necessary for 
the emission line to escape, requiring a significant contribution from other, probably fainter galaxies nearby}.
This scenario reflects LCDM  thinking, it postulates the existence of many unobserved faint galaxies and 
some a hoc tubes of transparency  between hydrogen clouds.
This by itself is already questionable, since in the local universe very few galaxies lie on the same sightline. 
More importantly, if the hydrogen is locked up in \mBDs, as we put forward, these faint galaxies and 
mythical clear sight-lines are not needed since then the universe is mainly empty and transparent.

\section{Conclusion and outlook}

Contrary to the current opinion, the galactic dark matter (DM) is likely of baryonic 
origin and observed already. It may seem ironic
that nature has sequestered most of the baryonic matter in nested clumps of
clumps, from planet mass to cluster and then to galactic scale. 
The structure predicted by the theory of gravitational hydrodynamics,
micro brown dwarfs (\mBDs, muBDs) grouped in Jeans clusters (JCs), is supported by 
an impressive body of observations:  quasar microlensing, planetary nebulae, 
15K cold dust temperatures, ``cirrus clouds'', extreme scattering events, parabolic events, direct 
observations and long duration radio events. This picture offers an explanation for paradoxes such as 
the Lyman-alpha forest, iron planet cores and young stars near the black hole in the center of the Galaxy.
Globular clusters seem to arise from JCs that are heated by tidal forces when passing nearby 
the galaxy's central black hole. 
Likewise, galaxy merging turns dark JCs into young globular clusters along the merging path~(\cite{NGS2009}), 
rather than producing tidal tails of old stars that have remained unobserved~(\cite{Duc}).

Fundamentally, our arguments show that evidence of the condensed hydrogen
objects constituting the baryonic dark matter have been accumulating for
many years, but have not been considered together. However, the
recent discovery that the Halo has the hydrogen temperature clinches the
case, and what had previously been called ``cirrus dust" must now be
understood as the missing hydrogen and other baryons. So we begin to hint
at other revisions to the fundamental theory needed to explain the origin,
nature, and properties of the baryonic dark matter.
 We summarize our findings in Table 2.
 
Also other observations can be easily explained in the GHD picture.
More binary Trans-Neptunian Objects than expected have been observed ~(\cite{Stephens2006});
This just matches with the the first stages of coagulation of \mBDs.
Brown dwarfs are known to exist but a description of their formation from hydrogen clouds 
is problematic ~(\cite{Burgasser2008});  Coagulation of \mBDs \ explains it immediately.
Radio astronomers may have found the first ever galaxy that is made almost entirely of dark matter.
In the Virgo cluster a ``dark galaxy" with hydrogen mass $\sim10^8M_\odot$ and dynamical mass 
$\sim10^{11}M_\odot$ exists with mass-to-light ratio of 500 times solar, 
which rotates in the same way as an ordinary galaxy, but without containing stars~(\cite{Minchin2005}); 
it may just have all its Jeans clusters still dark.
In the same M region of the Virgo cluster seen in the latter report, other
unexpected hydrogen clouds were found with hydrogen mass up to $10^8M_\odot$ and
dynamical mass up to $10^9M_\odot$ ~(\cite{Kent2010}). 
 
The direct search by Eros-I has concluded that the case of earth mass MACHOs is ruled out (\cite{Renault1998}).
However, for our \mBDs \  the Einstein radius that describes the lensing is comparable
with the physical radius of the lens object, which complicates the lensing.
This motivates new MACHO searches, which we plan to start in fall 2011,
benefiting from the improvement of CCD cameras
and taking into account the finite--source and finite--lens size effects.  
With the JCs in front of the Magellanic clouds and the Galactic bulge to be 
identified first by the Planck mission, or the ones detected already as ``cirrus clouds'' 
by DIRBE, BOOMERANG and WMAP, the task is reduced to searching lensing events by \mBDs 
\ in their JCs, a much more direct approach than the blind searches performed till now.

The cold dark matter paradigm seems to be left with a bleak face. In our picture its purported elementary 
particle has no reason to exist, because large scale structure formation can progress 
from gravitational hydrodynamics alone~(\cite{NGS2009}). 
Meanwhile the expected mass budget of the Galactic dark halo
is well estimated by our counting of cirrus clouds and our modeling of  \mBD \ mergings
as the observed radio events. 
For the extra-Galactic situation, these findings are in line with the optical depth in 
quasar micro-lensing being of order unity.
In line with this, the recent Xenon 100 and CDMS cold dark matter searches have ruled out all previous 
detection claims ~(\cite{Xenon100,CDMS2}). 
Massive neutrinos pose an alternative DM scenario~(\cite{NEPL2009}), which is cosmologically sound
~(\cite{NGS2009}). Detection of the 1.5 eV neutrino mass in Katrin 2015 would solve the notorious dark matter 
riddle in a dual manner, by dark baryons for the ``Oort'' galactic DM, clustered according to gravitational 
hydrodynamics, and by massive neutrinos for the ``Zwicky" cluster DM.

It would also be interesting to have more cases where the JCs can be resolved into their 
constituent \mBDs, such as it is possible in planetary nebulae. 
An interesting case is the star formation region in the Galaxy in the direction Crux, 
where individual Jeans clusters can be identified from their 14 K thermal emission in recent images~(\cite{ESA2009}).

Extreme reliance on linearized theories in astrophysics seems to have seriously misguided the
star formation theory. When direct simulations of gravitational perturbations in a gas cloud
showed collapse, the result was dismissed as unphysical and due to numerical instability 
~(\cite{Truelove1997}), and eliminated by Jeans filters. 
However, we consider the instability as physical, producing e.g. the discussed \mBDs, 
and the linearized theory as oversimplified. 
Interestingly, earth mass dark matter halos were discovered independently in a supercomputer 
analysis of the concordance model with neutralino dark matter,
predicting that the first structures to form have earth mass and size of the solar system
~(\cite{Diemand2005}).
Merging and infall of \mBDs \ then offers a new scenario for the formation of heavier objects,
from super-earths to heavy stars.
This work thus motivates the numerical study of the full nonlinear hydrodynamics 
of structure formation and the $N$-body dynamics of \mBD \ merging  as a precursor to
star formation and galaxy merging. 

\begin{center}     
\noindent     
\begin{tabular}{||l||r|r|r|r||rll|} 
 \hline \hline     
Observation&Jeans clusters& micro brown dwarfs& cosmol. signif.\\ \hline \hline
 galactic rotation curves & inferred & inferred & yes \\ \hline
 Tully-Fisher relation & infered & inferred& -- \\ \hline
 galaxy merging&yes & inferred & yes\\ \hline
 quasar microlensing& inferred& yes& yes\\ \hline
 planetary nebulae& inferred & yes& --  \\ \hline
 cold halo temperatures & -- & yes & -- \\ \hline
 cirrus clouds& yes & yes & yes \\ \hline
 extreme scattering events& inferred & yes& yes \\ \hline
 parabolic events & -- & yes& -- \\ \hline
 mysterious  radio events & yes & yes& yes \\ \hline
 star formation in Jeans cls & yes & inferred& - \\ \hline
 direct detection T-dwarf& -- & inferred & -- \\ \hline
 Ly-alpha forest& inferred& yes& inferred \\ \hline
Iron planet cores &  -- & inferred& -- \\ \hline
BH mass -- \# globulars & yes& inferred& --\\ \hline
visibility of early galaxies & inferred& yes & yes \\ \hline\hline
\end{tabular}     
\end{center}     
\vspace{6mm}   
Table 2. Observations discussed in the text, their relation to Jeans clusters and micro brown dwarfs
and their cosmological significance. 

\subsection{Acknowledgments}

We thank James Rich and Marcella Veneziani for discussion.

\appendix


\section{Properties of Jeans clusters}


Right after the transition of plasma to gas (decoupling of photons, recombination of electrons and protons)
there appears fragmentation at the Jeans length, creating JCs, and inside them at the viscous length,
creating \mBDs ~(\cite{NGS2009}).
Following Weinberg, we estimate the typical mass of a JC to be $6\cdot 10^5 M_\odot$.
Let $M_{jc}$ denote the typical mass of the JCs and $v_{jc}$ their velocity dispersion,
and $M_{bd}$ and $v_{bd}$ the ones of the \mBDs.
We introduce the dimensionless variables
$m_{6} = M_{jc}/6\cdot 10^5 M_\odot $,  $v_{200} = v_{jc}$/200 km s$^{-1}$,
$m_{bd} = M_{bd}/M_\oplus$,  $v_{30} = v_{bd}$/30 km s$^{-1}$;
these parameters are taken equal to unity in the main text.

For the distribution of \mBDs \ in a JC we consider the isothermal model 
$\rho_{bd}(r)=v_{bd}^2/2\pi G r^2$ where $v_{bd}=v_{bd}^{\it rot}/\sqrt{2}$ is the velocity dispersion 
in terms of the rotation speed. The number density is $n_{bd}=\rho_{bd}/M_{bd}$.
A JC has virial radius $R_{jc} = G M_{jc} /2v_{bd}^2= 4.4\cdot 10^{16}  (m_6 /v_{30}^{2})$ m
and contains $N_{bd}^{jc} =2.0\cdot 10^{11} (m_{6} /m_{bd})$ \mBDs.
The time for a \mBD \  to cross the JC is $R_{jc}/v_{bd}=47,000\,(m_{6}/v_{30}^3)$ yr, 
short enough to induce the isothermal distribution because of  Lynden Bell's ``violent relaxation''.
The average mass density is $3M_{jc}/4\pi R_{jc}^3=3.3\cdot 10^{-15}(v_{30}^6/m_6^2)$ kg/m$^3$.

Our Galaxy with mass $m_{12}10^{12}M_\odot$ has a number $N_{jc}=1.7\cdot 10^6 m_{12}/m_{6}$
of JCs in its halo, that also form an ideal gas with an isothermal distribution
$\rho_{jc}\equiv M_{jc}n_{jc}=v_{jc}^2/2\pi G r^2$, a shape central to the explanation of the flattening of rotation 
curves and the Tully-Fisher and Faber-Jackson relations~(\cite{NGS2009}).
The Galactic halo contains $3.3\cdot 10^{17} (m_{12}/m_6)$ \mBDs, only two orders of magnitude more 
than the number of earth mass objects estimated 
from simulation of LCDM cosmology ~(\cite{Diemand2005}) .

\section{Tadpole galaxy merging}

The beginning of the wake in Fig. 1 starts at $R_{gal}=4\cdot10^{21}$ m $=420,000$ lyr from the center.
Its width can be estimated at this point as $2R_{wake}=1.5\cdot 10^{19}$ m, and is kept fairly constant
over the entire length. For spherical symmetry the mass enclosed within $R_{gal}$ is 
$2v_{jc}^2R_{gal}/G=2.4\cdot10^{12} v_{200}^2M_\odot$, a reasonable value. We estimate
the number of JCs in this light path as the number in a cylinder of radius $R_{wake}$ through the center,
$N_{jc}^{wake}=\int_0^{R_{gal}}d z\int_0^{R_{wake}} \d r_\perp\,2\pi r_\perp n_{jc}(\sqrt{r_\perp^2+z^2})$,
yielding the estimate $N_{jc}^{wake}=5900\,v_{200}^2/m_6$, which is correct within a factor of a few,
given that some 11,000 were analyzed in Ref. ~(\cite{Fall2005}).

\section{Number of \mBDs \ in quasar micro lensing}

Let us estimate the number of \mBDs \  in the lens through which we observe the quasar Q0957A+B
discussed in the main text. We take Hubble constant $H=70$ km/(s Mpc).
The quasar has redshift $z_Q=1.43$ and angular distance $d_Q=5.6\cdot 10^{27}$ cm,
the lens has $z_L=0.355$ and $d_L=3.2\cdot 10^{27}$ cm. 
The most luminous part of the quasar is the inner ring of the accretion disk of $4\cdot 10^{16}$ cm in 
diameter and $10^{14}$ cm wide,  observed as angular diameters $1.5\,\mu$as and 3.7 nas, somewhat
distorted through shear in the lens.  
At the position of the lens this corresponds to a luminous ring of physical diameters  
$d_1=2.3\cdot10^{16}$ cm and $d_2=5.7\cdot 10^{13}$ cm, much less than the above JC radius,
but much more than the  \mBD \ radius of $2.6\cdot10^{11}$ cm discussed below.
The Einstein radius of an \mBD \  in the lens is 
$R_E=(2/c)(GM_{bd}d_L)^{1/2}=7.5\cdot 10^{13} m_{bd}^{1/2}$ cm, slightly larger than $d_2$,
so we are in the finite-source point-lens situation.

With the quasar observed in arm A under angle $\theta_A=5''$ with respect to the center of
the lensing galaxy and under $\theta_B=1''$ in the B arm,
the shortest physical distance of the light paths to the center of the lens galaxy is
$r_{A}=\theta_{A}d_L=7.8 \cdot10^{22}$ cm and  $r_{B}=1.6 \cdot10^{22}$ cm, respectively.
The average number of JCs that lie along the A sight line is $N_{A}=n_{2D}^{jc}(r_{A})\pi R_{jc}^2$, 
where $n_{2D}^{jc}(r)=v_{jc}^2/2GM_{jc}r$ is the projected isothermal number density. 
This brings $N_A=0.002\ m_6v_{200}^2/v_{30}^4$ and likewise $N_B=0.01\, m_6v_{200}^2/v_{30}^4$.
Being nearly of order unity, these numbers support the fact that microlensing events are observed in 
both arms and put forward that each of the two sight lines of the quasar pierces through one JC. 

Not having information about the distance of the sight lines to the centers of the JCs, we take for the
 $2D$ number density of  \mBDs \   the average value $n_{2D}=N_{bd}^{jc}/\pi R_{jc}^2$.
With active area $\pi d_1R_E$ this estimates that  $18,000\, v_{30}^4/m_6\sqrt{m_{bd}}$  \mBDs \ 
overlap the ring at any moment in time in either arm.
The optical depth, i. e.,  the number of \mBDs \ that lie inside a cylinder with radius $R_E$,
takes the typical value $n_{2D}\,\pi R_E^2=58\,v_{30}^4/m_6$. (These estimates do not
account for the shear). The lower values of 1.35 and 0.35 deduced from the signal analysis 
arise when the sight lines are near the border of the JCs, where the \mBD \ density is lower
than average.
All by all, the observations are consistent with isothermal modeling of \mBDs \ in JCs.

\section{MACHOs in front of the Magellanic clouds}

The Small and Large Magellanic Clouds (MCs) are satellites of the Milky Way at distance
$L=48.5$ kpc for the LMC, having $\nu\,10^{10}$ stars in their surface of 9+84 ${\rm deg}^2$ 
on the sky,  with $\nu=O(1)$, per unit solid angle equivalent to $1.1\nu\,10^8$/deg$^2$.
The angular surface of a JC at distance $xL$ (we take the typical JC half way, $x=\half$) is
$\Omega_{jc}=\pi R_{jc}^2/x^2L^2=3.7\cdot 10^{-5}  (m_{6}^2/v_{30}^4)\,{\rm deg}^2$,
so the average number of stars of the MCs covered by a JC is $3900\,(\nu m_{jc}^2/v_{30}^4)$.
 The number of JCs up to 
distance $L$ is $N_{jc}^L=2v_{jc}^2L/GM_{jc}=1.5\cdot 10^6 (v_{200}^2/m_{jc})$, so
in front of the MCs there are some 3400 $v_{200}^2/m_{6}$ JCs.
They should be identified from direct thermal emission
by the WMAP and Planck satellites, as foregrounds that have to be
subtracted in order to study the cosmic microwave background.

When an \mBD \  comes within the Einstein radius
$R_E=\sqrt{4x(1-x)GM_{bd}L}/c$ $=2.6\,10^9$ $\sqrt{m_{bd}}$ m to the sightline of a small star,
it acts as a lens and enhances the intensity during a time
$t_E=R_E/v_{jc}=3.6\sqrt{m_{bd}}/v_{200}$ hrs.
The \mBDs  \  that lens a given star within a period $\Delta t$,
lie in a strip on the sky with surface $\Delta O=2 R_Ev_{jc}\Delta t$, which
embodies $N_{bd}^{jc} \Delta O/\pi R_{jc}^2$ \mBDs. This leads to a lensing event rate of
$1.1\, (v_{200} v_{30}^4/\sqrt{m_{bd} }m_{6})/$yr for each star covered by a JC
and monitoring this frequency allows to estimate $m_{jc}$.
With 3400  $m_6^2 \nu/v_{30}^4$ stars of the MCs per JC, this leads 
4100 $(m_6 \nu v_{200}/\sqrt{m_{bd}})$ of Macho events per yr per JC
and in total $1.4\cdot 10^7\,\nu v_{200}^3/\sqrt{m_{bd}}$ events per year
in front of the MCs. 

In the Eros--I search some $2\cdot10^6$ stars were monitored,
so we would expect some 2800 lensing events per year, while none was discovered.
It was realized that the stellar radii are at least comparable to the 
Einstein radius (finite source); no lensing detection is made when the stellar radius 
exceeds 16 solar radii, which excludes a sigificant part of the studied stars. 
However, also the deflector is extended (finite lens), because the Einstein radius of the \mBD \ 
just coincides with our estimate for its physical radius. This may lead to obscuration, and on top of this, 
there will be refraction effects from the atmosphere of the \mBD: 
Lensing of earth mass \mBDs \ towards the MCs appears to be a subtle case, that calls for more study.

Each JC has $N_{bd}^{jc}$ \mBDs \   with Einstein area 
$\Omega_E=\pi R_E^2/x^2L^2=1.2\,10^{-19}  m_{bd}\,{\rm deg}^2$.
This brings a coverage fraction or ``optical depth" equal to
$\tau=N_{bd}^{jc}\Omega_E N_{jc}^L/8\pi=v_{jc}^2/c^2=4.9\,10^{-7}v_{200}^2$, as is well known.
But JCs themselves have a much larger optical depth, 
$ \tau=N_{bd}^{jc}R_E^2/ R_{jc}^2=6.7\cdot 10^{-4}(v_{30}^4/m_{6})$,
so it is of course advantageous  to identify the JCs first and monitor the light from stars behind them.

\section{Composition of cirrus clouds}

Fig. 3 of the cirrus clouds shows mass concentrations with kJy emission~(\cite{Veneziani2010}).
Let us assume that this DIRBE signal at $\nu=1.25$ THz stems from thermal emission at the \mBD \
surface with temperature $15$ K,  with the Planck function equal to 
$I_\nu=4\pi  \hbar  \nu^3/[\exp(2\pi \hbar \nu/k_B T) - 1]c^2=54$ GJy.
We first investigate whether it may arise from individual \mBDs.
The distance to the nearest \mBD \ can be estimated if the Sun is inside a local Jeans cluster. 
The typical number density $3M_{jc}/4\pi R_{jc}^3M_{bd}$ leads to a typical 
distance $d_1=1.2\cdot 10^{15} (m_6^{2/3} m_{bd}^{1/3}/v_{30}^{2})$ cm, or 82 AU,
which is well before the Oort cloud, believed to start at 2000 AU. 
With the \mBD \ radius taken from previous section, we get an intensity $J=I_\nu R_{bd}^2/d_1^{2}=
2.4\, (t_d^2 v_{30}^6 /m_6^{4/3} m_{bd}^{2/3})$ kJy, the observed scale.
But the angular diameter of an \mBD \ at this distance is $0.02^\circ$, 
much smaller than the structures in Fig. 3. 

The signal is therefore more likely arising from JCs. Their local number density is isothermal,
$n=v_{jc}^2/2\pi G M_{jc}d_{SagA^\ast}^2$, involving the distance 25,000 lyr to Sag. A$^\ast$,
and yields the typical distance to next JC as $d_2=n^{-1/3}=8.9\cdot 10^{18} (m_6^{1/3}/v_{200}^{2/3})$ m
or 940 lyr. The $2\cdot 10^{11}$  \mBDs \ in next JC bring together also a kJy signal, 
$J=N^{jc}_{bd} I_\nu R_{bd}^2/d_2^{2} = 0.92 \,(t_d^2 m_6^{1/3} v_{200}^{4/3} v_{30}^2/m_{bd}) $ kJy.
At this distance the JC diameter $2R_{jc}$ appears at an angle of
$\theta=2R_{jc}/d_2=(m_6^{2/3}v_{200}^{2/3}/ v_{30}^2)\,0.57^\circ$, in good agreement 
with the largest structures of Fig. 3.
The angular surface of the JC is $\Omega=\pi \theta^2/4$. The intensity per unit solid angle is 
$\d J/\d \Omega\approx J/\Omega=I_\nu N_{bd}^{jc}R_{bd}^2/\pi R_{jc}^2$sr $=
11.7 (t_d^2 v_{30}^6/m_6 m_{bd})$ MJy/sr.
This is essentially the right order of magnitude, since several structures in fig. 3 have the maximal intensity 
of 2 MJy/sr, which is achieved for JCs with proper values of $t_d$ and $m_6$.

Let us estimate the number of structures in Fig. 3.
Arising from JCs with equal intensity at varying distances, 
they form clumped structures with diameter of 1$^\circ$ down to, say,
0.25$^\circ$, corresponding to distances $d_1=5.1\cdot10^{18}  m_6/v_{30}^2$ m and
$d_1'=2.0\cdot10^{19}  m_6/v_{30}^2$ m, respectively.
The number of JCs in a spherical shell between them is $N_{4\pi}=2v_{jc}^2(d_1'-d_1)/GM_{bd}
=15,000v_{200}^2/v_{30}^2$. In a DIRBE area one gets as number of JCs
$N_{4\pi}15^\circ\times12^\circ/4\pi {\rm rad}^2=67v_{200}^2/v_{30}^2$, which is a fair estimate
for the number of structures in Fig. 3.

We can therefore conclude that these concentrations likely arise from JCs
that constitute the full Galactic dark matter.

\section{Rate of \mBD \ merging}

 We attribute the observed radio events to merging of \mBDs \ in JCs that constitute 
 the BDM of the Galaxy. Let us present a statistical estimate for its frequency.
{}From the event duration $t_{ev}$ the effective \mBD \  radius is estimated as 
$R_{bd} = v_{bd} \,t_{ev}=2.6\cdot 10^{11}t_d v_{30}$ cm.
The typical JC has $N_{cell}^{jc}=(R_{jc}/R_{bd})^3$ cells of linear size $R_{bd}$.
At a given moment in time its typical number of events, i. e., the number of  \mBD \  pairs
that occupy the same cell, is of order unity,
$\half N_{bd} ^{jc}(N_{bd} ^{jc}-1)/N_{cell}^{jc} = 0.50\, t_d^3 v_{30}^9/ m_{6}m_{bd}^2$.
The rate of mergings is this number divided by $t_e$, viz.
$3.1\cdot10^{8}(m_{12}t_d^2 v_{30}^9/ m_{6}^2m_{bd}^2) /$yr, corresponding to
$7400\,(m_{12}t_d^2 v_{30}^9/ m_{6}^2m_{bd}^2)/$deg$^2$yr.
This estimates fairly well the observed rate $\sim1000/{\rm deg}^2$yr and the extraplolated full sky rate 
$4.1\,10^7$/yr is equivalent to $6\cdot10^{17}$ merging events in the Hubble period of 14 Gyr,
which amounts to two merging events per \mBD. Given that all large objects, like the Sun,
arise from merged \mBDs, there should still be many original \mBDs, and this is observed
in Fig. 2  of Helix.

The amount of energy available in the merging process is large, of order $GM_{bd}^2/R_{bd}$,
some $10^5$ J/kg.
Assuming this to be emitted at distance of 3 kpc during the event time of 1 day in a frequency
band of 10 GHz gives a bright radio event of 1 Jy; such events are observed,
see table 2 of ~(\cite{Ofek2010}.

\section{Lyman-alpha forest}

Let us consider a quasar of mass $M_q=m_{9}10^{9}M_\odot$ at redshift $z_q=3$.
It has a Schwarzschild radius $R_q=2GM_q/c^2=3.0\cdot10^{12}m_9$ m and angular distance
$d_A(z_q)=5.1\cdot10^{25}$ m.
A JC, at redshift $z$, is much larger, the number of its \mBDs \  in front of the quasar 
$N_{bd}^q(z)=N_{bd}^{jc}(R_q/d_A(z_q))^2(d_A(z)/R_{jc})^2$ equals
$890 (m_9^2 v_{30}^4/m_6 m_{bd})d_A^2(z)/d_A^2(z_q)$. 

JCs intersected by the quasar sight line lie in a cylinder of radius $R_{jc}$, so their mass can be estimated 
as $0.02\rho_c\cdot\pi R_{jc}^2\cdot c/H_0$, where the factors describe the average cosmic baryon density 
not in X-ray gas, the surface area and typical quasar distance, respectively. 
This corresponds to 0.0001 JCs. But the typical density in filaments between cosmic voids is $5-10$ times 
the average galactic density of 0.2 $M_\odot$/pc$^3$, some $10^7\rho_c$, 
so the clumpedness of matter can explain that indeed \mBDs \ cause a narrow absorption line per JC.

\newcommand{\asas}{Astron. \& Astrophys}

\end{document}